# A European research roadmap for optimizing societal impact of big data on environment and energy efficiency

Martí Cuquet, Anna Fensel
Semantic Technology Institute
Universität Innsbruck
Innsbruck, Austria
{marti.cuquet, anna.fensel}@sti2.at

Lorenzo Bigagli
National Research Council of Italy
Firenze, Italy
lorenzo.bigagli@cnr.it

*Abstract*— We present a roadmap to guide European research efforts towards a socially responsible big data economy that maximizes the positive impact of big data in environment and energy efficiency. The goal of the roadmap is to allow stakeholders and the big data community to identify and meet big data challenges, and to proceed with a shared understanding of the societal impact, positive and negative externalities, and concrete problems worth investigating. It builds upon a case study focused on the impact of big data practices in the context of Earth Observation that reveals both positive and negative effects in the areas of economy, society and ethics, legal frameworks and political issues. The roadmap identifies European technical and non-technical priorities in research and innovation to be addressed in the upcoming five years in order to deliver societal impact, develop skills and contribute to standardization.

## I. INTRODUCTION

ICT and particularly Internet of Things (IoT) solutions are now being intensively applied when addressing energy efficiency challenges: they range from already traditional systems performing automation to the ones addressing more innovative aspects, for example, behavioral change [1]. What is common in such systems is that the use of IoT solutions leads directly to the generation of Big Data of various kinds: building and transport data, environment data, context data, user data [2].

Consequently, Big Data, commonly characterized by its volume, velocity, variety, variability, and veracity [3], create a large societal impact in general, and in particular in the energy efficiency, sustainability and environment sectors. Issues to address comprise ethical considerations, data licensing, user data management, privacy and security, and open data publication among others. The use and misuse of IoT-generated Big Data are on the rise, witnessed for example in the recent news headlines about ransomware, i.e. electrical appliances controlled from outside and not by their owners. A growing number of projects and initiatives are investigating the effects of the publication of IoT energy and environment data as open data. As an example, the OpenFridge project has been producing and publishing refrigerator data as open data, and studying the user perception and possible uses of such data [4].

To ensure an effective data value chain, it is highly important to investigate interdependencies between different data sets. For example, environment data is a natural crucial enabler for energy efficiency aims. On the one hand, it provides information of the status quo of our environment, on the basis of which ICT tools for energy efficiency can be built. On the other hand, the same kind of data are to be used later in the evaluations and accountability of the taken energy efficiency measures.

The goal of the present roadmap is to provide incremental steps necessary to maximize the societal impact of big data research on environment, and its related influence on energy efficiency and sustainability. It is aligned with a broader, multi-sectorial research [5] and policy [6] roadmap that has resulted from the Big data roadmap for cross-disciplinarY community for addressing societal Externalities (BYTE) Project, a three-years European project involving the research, industry and civil society community, and with the Big Data Value Strategic Research and Innovation Agenda (BDV SRIA) that defines the overall goals and technical and non-technical priorities for the European Public Private Partnership on Big Data Value [7]. A community of research, civil society organizations and industry partners, has been built around the roadmap with the purpose of implementing its recommendations and best practices and further develop and update it. The community has initially focused on the environment, healthcare and smart city sectors.

This paper is organized as follows. Section II presents the scope of the roadmap (research areas covered, time span and desired impact) and the methodologies used in the environment case study and roadmapping process. Section III summarizes the case study that forms the base of the requirements analysis for the roadmap, and Section IV outlines the research priorities, timeline and expected impact. We conclude in Section V with some remarks on the expected societal benefits.

## II. METHODOLOGY

### A. Roadmap scope

In this paper, we cover research and innovation in five technical areas—data management, data processing, data analysis, data protection, and data visualization—and present which topics have the highest priority to deliver a societal impact to the energy and environmental sector in the upcoming 5 years, as well as the required skills development and standardization efforts. Phaal, Farrukh and Probert [8] propose a T-Plan fast-start approach to technology roadmapping that is primarily developed for use from a company perspective, but can be customized for a multi-organizational use of a group of stakeholders, and it has been explicitly done so in the context of disruptive technological trends.

The roadmap has been developed following such multilayered approach [8], as sketched in Table 1. We defined four top layers, sometimes labeled as *know-why*, that encapsulate the organizational purpose and correspond to the externalities that the roadmap is intended to impact and potentiate (if positive) or diminish (if negative). The externalities are arranged in four areas and 18 coarse-grained externalities. In addition to these purpose layers, we also considered how this research impacts the energy efficiency and environment sector. This is part of a larger effort to include up to 18 sectorial layers [5] that represent the society pull in the roadmap. We further defined six bottom layers, also known as *know-how*, corresponding to the five-technical and the non-technical research and innovation areas, or resources, that are to be addressed to meet the demands of the top layers, and that encode the technology push. Of these areas, topics in data analysis were not found a priority for a societal impact in the environment sector. Finally, the middle layers of the roadmap connect the purpose with the resources to deliver benefits to stakeholders, i.e. represents the *know-what*. This includes the skills development, standardization efforts and societal impact that the research and innovation actions contribute to.

**Table 1. Layer structure of the roadmap.**

| | |
|---|---|
| Economic externalities | Purpose (know-why) |
| Social and ethical externalities | |
| Legal externalities | |
| Political externalities | |
| Societal impact | Delivery (know-what) |
| Skills development | |
| Standardization | |
| Data management | Resources (know-how) |
| Data processing | |
| Data analytics | |
| Data protection | |
| Data visualization | |
| Non-technical priorities | |

*B. Case study methodology*

The roadmap is the culmination of a series of case studies, analysis, expert focus groups and workshops conducted within the BYTE project [9] that resulted in an identification of the societal impacts of big data in seven European sectors: crisis informatics [10], culture, energy [11], environment, healthcare, smart cities and transport.

A total of 73 societal externalities related to big data usage were identified [12]. For this study, we extended the economical concept of externality to include not only economical but also social, ethical, legal and political benefits or costs on third-parties arising from big data practices, including potential activities, opportunities and risks. These externalities were classified by the pairs of stakeholders involved (public sector, private sector and citizens) and their main topic (business models, data sources and open data, policies and legal issues, social and ethical issues, and technologies and infrastructures).

Following qualitative research [13] and case study methodologies [14], the case studies findings were supported by multiple sources of evidence. To compile information about the main data sources, their uses and data flows and the challenges faced by the community in the environment case study, conducted in the context of a global effort for Earth Observation, we performed six semi-structured interviews with senior data scientists and IT engineers, participated in the 4th GEOSS Science & Technology Stakeholder Workshop to gather first-hand input from the community, and held a focus group in April 2015 in Vienna with 11 environment experts from academia and industry, selected to ensure the participation of individuals with expertise in environmental data, technology, computer science, standardization, the Space sector, as well as privacy and data protection, open data policies, relevant policy issues such as funding and innovation [15]. The data analysis was performed via the interviews and focus group transcripts, data coding, and themes identification and analysis [13], [16], [17].

*C. Roadmapping*

The 73 externalities were simplified to 18, and grouped in four main areas in the context of a horizontal analysis across all seven sectors [18]:

**Economic externalities:** improved efficiency, innovation, changing business models, employment, and dependency on public funding.

**Social and ethical externalities:** improved efficiency and innovation, improved awareness and decision-making, participation, equality, discrimination, and trust.

**Legal externalities:** data protection and privacy, intellectual property rights, and liability and accountability.

**Political externalities:** private vs. public and non-profit sector, losing control to actors abroad, improved decision-making and participation, and political abuse and surveillance.

Relevant research and innovation topics were then identified, mapped to their relevance to address societal externalities, and analyzed how they may impact society and contribute to standardization and skills development. Such relevance and mapping was assessed by a review of the case study reports [15] and complemented with an analysis of big data initiatives and external studies [7], [19]–[21] to include each significant contribution to the roadmap.

These results were validated in a dedicated research roadmapping workshop collocated with the European Data Forum 2016 in Eindhoven, with the assistance of 26 participants from academia (11), SMEs (8), large companies (3), public organizations (3) and certification bodies (1), coming from 11 European countries (Austria, Belgium, Germany, Hungary, Ireland, Italy, the Netherlands, Norway,

Spain, Sweden, and the United Kingdom). The workshop also contributed to the prioritization and time-alignment of the topics. A second workshop with the big data community, collocated with the Big Data Value Association Summit 2016 in València, validated the specific outcomes for the environment sector.

### III. ENVIRONMENT CASE STUDY

The environment case study was conducted in the context of the Group on Earth Observation (GEO), a global-scale initiative for Earth Observation, involving more than 100 national governments and in excess of 100 NGOs, for better understanding and controlling the environment, to benefit society through better-informed decision-making. GEO has been driving the interoperability of many thousands of individual Space-based, airborne and in situ stations for Earth Observations around the world for more than a decade. In this section, we focus on GEO's data sources and uses, and the societal externalities present in the case study. A more detailed analysis of its technological aspects and data flows can be found in [15].

*A. Data sources*

The case study revealed a large number of heterogeneous environmental data sources gathered from hundreds of countries, several thousand locations, ships, aircraft, land vehicles, satellites. The interlinking of data is seen as a source of new data itself providing new insights, especially when such linking is done with e.g. non-authoritative, unstructured data from social media. Data sources revealed by the interviewees are those with a Space (satellite data), in-situ component (rain gauges, buoys), or service component (models), or come from cadaster and utilities infrastructure data, open and public sector information data, archives, historical and archeological data, government agencies, time series of Earth Observation products (climate or weather data), linked data (e.g. that of the European Environment Agency, accessible through a SPARQL interface), web and social media (e.g. Twitter indications of earthquake extent), volunteered data from citizens and data from Internet of People (e.g. health monitoring) and Internet of Things.

*B. Data uses*

In the environment case study, big data shows up more as an incremental change rather than as a major shift. Data practices show up in all steps of the Big Data Value Chain [22]. The data acquisition comes mainly from data streams of national and international Space agencies and the remote sensing industry, and sensor networks from Government Environmental Agencies. Data analysis includes that performed by the community (e.g. combining and report on data for a municipality), cross-sectorial data analysis, information extraction and stream mining, and linked data and semantic analysis. The main activities within the data curation step are those dealing with interoperability, improving the quality, reliability, management and accessibility of data of importance to all fields of science and technology, and providing the required infrastructure to support open access and legal interoperability. Finally, on the data usage end, we found prediction of crisis and impact forecasting, data as a support for decision-making in huge processing demands caused by crisis, civil protection agencies and disaster management, in-use analytics, domain-specific usage (e.g. in agriculture, tourism or food industry), control (traffic, antiterrorism, policy enforcement and global monitoring of international agreements), and modeling and simulation.

The main technical challenges observed were in resolution, data discovery and integration, transformation into actionable data, quality and trust of data sources, sustainability to provide continuous and long-term access to data, better interpretation of models, and finally lack of standards and the existence of industrial competitors that use standard violations to strengthen their position.

*C. Societal externalities of big data*

To assist the development of a research and innovation agenda with a societal impact, the effect of big data practices in the environment case study was examined. The outcomes show that most practices may produce both positive and negative impacts, and it is thus necessary to provide the necessary means for capturing positive externalities (e.g. providing social benefits) while reducing the negative ones (e.g. data misuse). We have classified such externalities as economic, social and ethical, legal, and political.

It was broadly recognized that providing reliable environmental data has a strong impact on economies. This is especially relevant in Europe due to its leading role in the Earth Observation industry. Negative economic externalities of big data usage also showed up as a threat to traditional services (e.g. weather forecasting) and the potential increase of market inequalities (open access policies by public bodies that put the big private players at a competitive advantage), which may be alleviated by niche opportunities for new or small companies.

Positive social and ethical impacts revolved around improved governance of environmental challenges and increase social awareness and participation. One example of positive externality is improved decision-making for sustainable development and disaster risk management, leading to progress to more sustainability, environmental safety and reduced disaster risk. Energy efficiency benefits also showed up in other related case studies [9], especially in sectors closer to the environment one such as the oil & gas and the smart city case studies. For example, big data was found to help reduce environmental impacts by the early detections of incidents and by monitoring equipment condition [11] and to better resource efficiency through targeted services in the smart city.

Positive benefits were partially hindered by the fear of data abuse and privacy violation. Excessive trust in data-intensive applications was also highlighted as a negative implication that encourages overlooking fundamental qualitative aspects.

Most legal externalities showed up as negative effects arising from the shortcoming of current legal frameworks on IPR, privacy, etc., which were amplified by the myriad of different legislations. The uptake of data-intensive applications was seen however as a positive effect prompting better informed and more precise legislation.

Finally, and from a political point of view, big data usage is expected to make political decisions more transparent and accountable. This comes at a risk of too much dependence on external sources, especially those from big players, who were perceived as imposing their commercial products as de facto standards and its associated lock-in.

## IV. RESEARCH PRIORITIES AND ROADMAP

In this section, we present the research priorities to deliver a societal impact in the environment sector. The timeline to address these topics, together with their priority level as assessed by the community of stakeholders, are shown in Fig. 1, and their expected societal impact in Fig. 2. The remaining of the section offers a description of such topics and impact.

**Data management** priorities are measuring and assuring data quality [23], research into data provenance, control and IPR to minimize threats to intellectual property rights (including scholars' rights and contributions) [23], [24] and the development of the data-as-a-service model and paradigm to exploit new opportunities for economic growth (new products and services based on open access to big data) [12]. They also include scalable data access mechanisms [25]. It has to contribute to fix, on a technical level, the current lack of norms for data storage, processing and use. New models should be encouraged that diminish inequalities to data access between big data players and the rest. In the related smart city sector, increasing semantic interoperability for urban multimodal transportation, sensor, social media and user-generated data from e.g. citizens' smartphones would among other benefits enhance the energy efficiency of the city [23], [25].

**Data processing** priorities revolve around techniques and tools for processing real-time heterogeneous data that help gather public insight by identifying environmental trends and statistics [26], and open decentralized architectures that diminish storage costs and decrease the dependency on external data sources, platforms and services [24]. With the exponential increase of data storage needs, developing energy efficient mechanisms for storage and processing also came up as transversal research priority among all studied sectors [24].

In the area of **data protection**, the main priority is the development of pattern hiding mechanisms and privacy preserving mining algorithms to avoid discriminatory practices and targeted advertising [12], [19]. In **data visualization**, end user visualization and analytics need to ensure that manipulation of visual representations of data is avoided [23]. Moreover, new visualization for geospatial data should help understand and manage environmental data and enable data-driven policy-making [27]. This has repercussions to sectors other than environment that will benefit from easy access to such reports.

Finally, a number of **non-technical priorities** have also been identified. First is to establish and increase trust via better transparency and accountability of the public sector [25]. Also, the increasing awareness about privacy violations and ethical issues of big data can be met by developments in privacy-by-design, security-by-design, anti-discrimination-by-design frameworks [19], [24], [25]. This will decrease the public reluctance to provide information (especially personal data), threats to data protection and personal privacy, and contribute to overcome the reduced innovation due to restrictive legislation [28]. Ethical issues should be investigated around sample bias and "sabotage" data practices, e.g. in social media fraud profiles willingly misinforming and possibly creating false data that affect the overall picture [27], [29]. New business models with closer linkages between research and innovation to capture opportunities for economic growth based on open access to big data have to be developed [7], [19], [23]. Examples put forward by the case study were the use of sea data for fishing purposes and weather data in the tourism industry. These new models may contribute to diminish the dominance of big market players. The environment sector shows great potential to encourage citizen research, which may take the form of crowd-computing, pervasive-computing, crowd-sourcing or independent research using open data and tools to e.g. increase data accuracy [25] or scale data curation [23]. To enable it, such tools have to be developed. This would increase citizen participation, produce safe and environment-friendly operations, deliver better models, measures and test about preparedness and resilience of communities, as well as of human behavior under crisis, and generally enhance quality of life. Furthermore, a strong participation of the public sector can help make data and services from the environment sector to become public goods available to all.

The need for data skills have been repeatedly put forward across all sectors [19], [20], [30], [31]. Such skill profiles can be broadly classified in three categories: data scientists or deep analytical talents to analyze the data, data-intensive business experts or data-savvy managers to effectively consume it, and data-intensive engineers or supporting technology personnel [7], [30]. To deliver the outlined societal benefits, environment sector stakeholders agreed that the sector has a mainly a strong requirement of data scientists to cover new data-driven employment offerings that will result from the challenge of traditional non-digital services, such as traditional weather forecasting. Furthermore, technology and data standards need to be developed to enhance data-driven R&D. Finally, skill development should also be aimed at diminishing inequalities to data access and the data divide [12].

## V. CONCLUDING REMARKS

This paper has presented a research and innovation roadmap to address the societal externalities of the use big data in the environment sector. The environment data have a direct impact on energy efficiency, as they are usable in scenarios such as climate change observation and prediction, and analysis of the smart cities' structures and their development.

The roadmap is applied to a case study performed in the context of the Group on Earth Observation (GEO), a multi-sectorial, global-scale initiative for Earth Observation, involving more than 100 national governments and over 100 Participating Organizations. The case study was one of seven conducted in the framework of the BYTE project, a European-wide project to help Europe capture a greater share of the big data market by using big data responsibly.

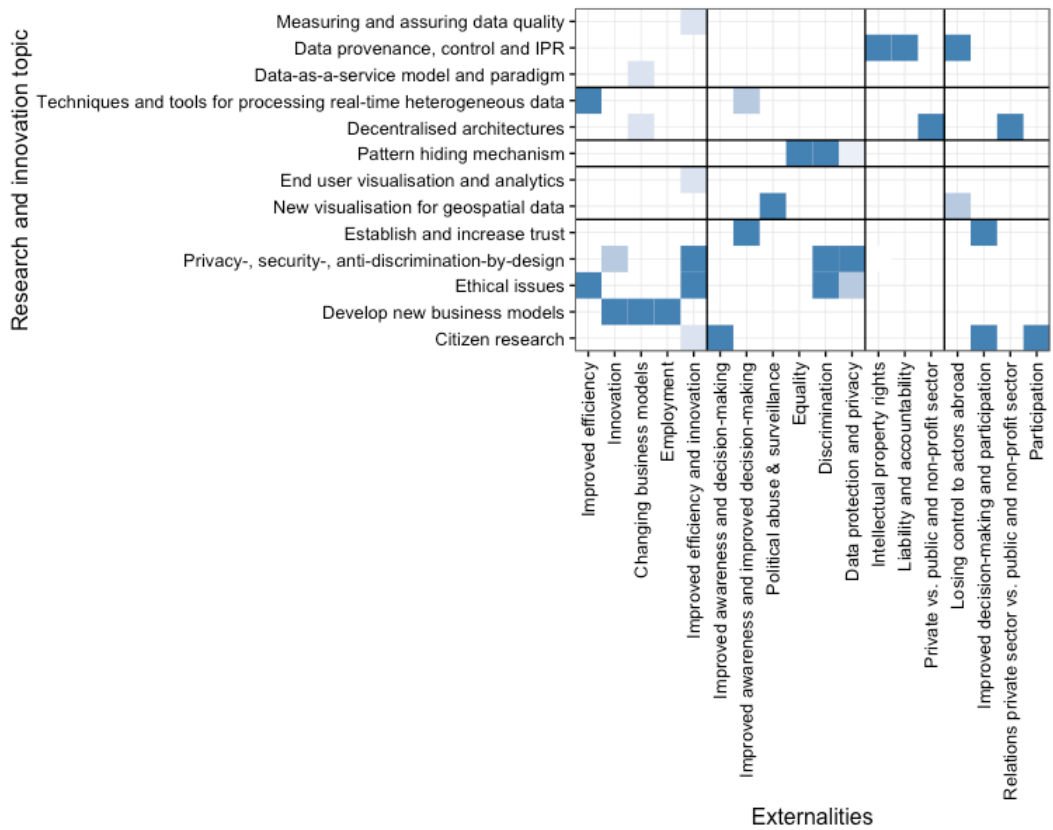

Fig. 1. Research and innovation topics with an expected impact on the environment sector. Topics have been grouped in areas (top to bottom): data management, data processing, data protection, data visualization and non-technical priorities. Relevance has been assessed by workshop participants (from dark to lighter blue): top priority if all or almost all stakeholders agreed the topic to be of high priority; high priority if it was generally considered to be of high priority; medium priority if it was generally considered to be of medium priority; low priority otherwise.

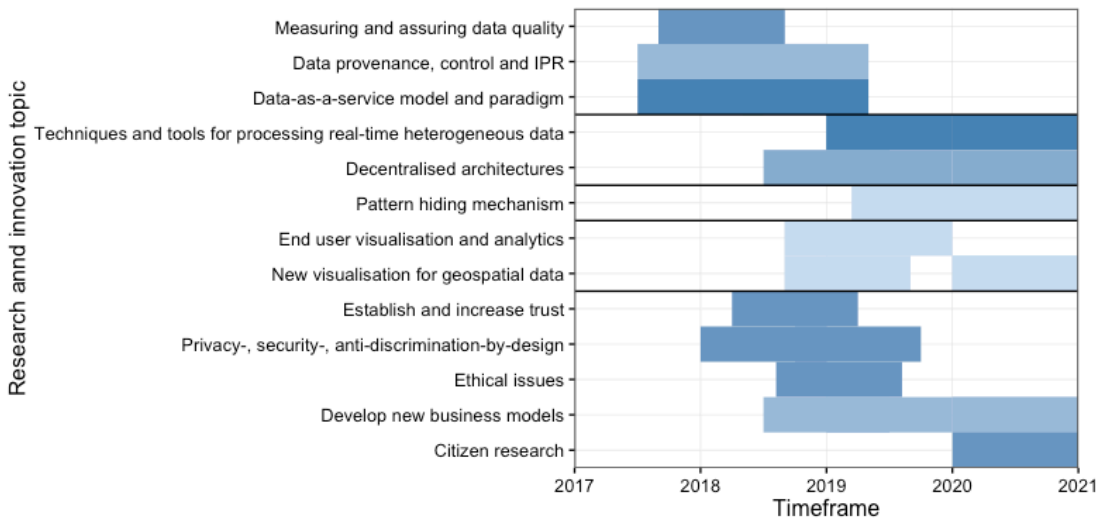

Fig. 2. Expected impact of the research and innovation topics in the environment sector. Relevance has been normalized by externality: darkest blue corresponds to the most relevant research in the externality. Research topics are grouped in the following areas (top to bottom): data management, data processing, data protection, data visualization, non-technical priorities. Externalities are grouped in four areas (left to right): economic, social and ethical, legal, and political.

The implementation of this roadmap is expected to deliver key social benefits in the areas of decision-making, data-driven innovations and novel business models, environmental benefits, and increased citizen participation, transparency and trust [9]. Improved decision-making comes especially from the use of data analytics for large volumes of environmental data to impact legislation and policies, as well as more efficiency in resource allocation. This would also have an impact on public health. Linking, integrating and processing heterogeneous data is expected to deliver new insights and data-driven innovations. This can be exemplified from the case study with the use of high-resolution satellite data and building information models to assess vulnerabilities. Big data usage is also envisaged to result in safer and more environmentally friendly operations, and enhance public health through better handling of environmental predictors (e.g. pollution levels). Finally, the environmental sector offers a highly favorable playground for crowdsourcing of data collection, and climate data portals are an opportunity to make policy-makers accountable, in particular for the implementation of energy efficiency and climate change objectives.

ACKNOWLEDGMENT

We thank the participants of the BYTE project focus groups and workshops for their insightful contributions.

This work has been partially funded by the European Commission through the BYTE project [FP7 GA 619551].